\documentclass[twocolumn,showpacs,prb,superscriptaddress,a4paper,floatfix]{revtex4}
\usepackage{graphicx}
\usepackage{amsmath}
\usepackage{graphicx}
\usepackage{dcolumn}
\usepackage{bm}

\begin{document}

\title{Decoherence by a spin thermal bath: Role of the spin-spin interactions and initial state of the bath%
}
\author{Shengjun Yuan}
\affiliation{Department of Applied Physics, Zernike Institute for Advanced Materials,
University of Groningen, Nijenborgh 4, NL-9747 AG Groningen, The Netherlands}
\author{Mikhail I. Katsnelson}
\affiliation{Institute of Molecules and Materials, Radboud University of Nijmegen, 6525
ED Nijmegen, The Netherlands}
\author{Hans De Raedt}
\affiliation{Department of Applied Physics, Zernike Institute for Advanced Materials,
University of Groningen, Nijenborgh 4, NL-9747 AG Groningen, The Netherlands}
\date{\today }

\begin{abstract}
We study the decoherence of two coupled spins that interact with a spin-bath
environment. It is shown that the connectivity and the coupling strength
between the spins in the environment are of crucial importance for the
decoherence of the central system. For the anisotropic spin-bath, changing
the connectivity or coupling strenghts changes the decoherence of the
central system from Gaussian to exponential decay law.
The initial state of the environment is shown to affect
the decoherence process in a qualitatively significant manner.
\end{abstract}

\maketitle





\section{Introduction}


Understanding the decoherence in quantum spin systems is a subject of
numerous works (for reviews, see Refs~\cite{stamp,ZhangW2007}). The issue
seems to be very complicated and despite many efforts, even some basic
questions about character of the decoherence process are unsolved yet. Due
to the interactions with and between the spin of the bath, an analytical
treatment can be carried out in exceptional cases, even if the central
systems contains one spin only. Recent work suggests that the internal
dynamics of the environment can be crucial to the decoherence of the central
system~\cite{Dawson2005,Rossini2007,Tessieri2003,Camalet2007,YuanXZ2005,
YuanXZ2007, Wezel2005,Bhaktavatsala2007,ourPRE,Relano2007,ZhangW2006,
JETPLett,Yuan2007}. In this paper, we present results of extensive
simulation work of a two-spin system interacting with a spin-bath
environment and show that the decoherence of the two-spin system can exhibit
different behavior, depending on the characteristics of the coupling with
the environment, the internal dynamics and the initial state of the latter.
We also provide a simple physical picture to understand this behavior.

In general, the behavior of an open quantum system crucially depends on the
ratio of typical energy differences of the central system $\delta E_c$ and
the energy $E_{ce}$ which characterizes the interaction of the central
system with the environment. The case $\delta E_c \ll E_{ce}$ has been
studied extensively in relation to the ``Schr\"{o}dinger cat'' problem and
the physics is quite clear~ \cite{zeh,zurek}: As a result of time evolution,
the central system passes to one of the ``pointer states''~\cite{zurek}
which, in this case, are the eigenstates of the interaction Hamiltonian $%
H_{ce}$. The opposite case, $\delta E_c \gg E_{ce}$ is less well understood.
There is a conjecture that in this case the pointer states should be
eigenstates of the Hamiltonian $H_c$ of the central system but this has been
proven for a very simple model only~\cite{paz}. On the other hand, this case
is of primary interest if, say, the central system consists of electron
spins whereas the environment are nuclear spins, for instance if one
considers the possibility of quantum computation using molecular magnets~%
\cite{m1,m2}.

\section{Model}

We consider a generic quantum spin model described by the Hamiltonian $%
H=H_{c}+H_{ce}+H_{e}$ where $H_{c}=-J\mathbf{S}_{1}\cdot \mathbf{S}_{2}$ is
the Hamiltonian of the central system and the Hamiltonians of the
environment and the interaction of the central system with the environment
are given by
\begin{eqnarray}
H_{e} &=&-\sum_{i=1}^{N-1}\sum_{j=i+1}^{N}\sum_{\alpha }\Omega
_{i,j}^{(\alpha )}I_{i}^{\alpha }I_{j}^{\alpha },  \notag \\
H_{ce} &=&-\sum_{i=1}^{2}\sum_{j=1}^{N}\sum_{\alpha }\Delta _{i,j}^{(\alpha
)}S_{i}^{\alpha }I_{j}^{\alpha },  \label{HAM}
\end{eqnarray}%
respectively. The exchange integrals $J$ and $\Omega _{i,j}^{(\alpha )}$
determine the strength of the interaction between spins $\mathbf{S}%
_{n}=(S_{n}^{x},S_{n}^{y},S_{n}^{z})$ of the central system, and the spins $%
\mathbf{I}_{n}=(I_{n}^{x},I_{n}^{y},I_{n}^{z})$ in the environment,
respectively. The exchange integrals $\Delta _{i,j}^{(\alpha )}$ control the
interaction of the central system with its environment. In Eq.~(\ref{HAM}),
the sum over $\alpha $ runs over the $x$, $y$ and $z$ components of spin-$%
1/2 $ operators $\mathbf{S}$ and $\mathbf{I}$. In the sequel, we will use
the term \textquotedblleft Heisenberg-like\textquotedblright\ $H_{e}$ ($%
H_{ce}$) to indicate that each $\Omega _{i,j}^{(\alpha )}$ ($\Delta
_{i,j}^{(\alpha )} $) is a uniform random number in the range $[-\Omega
|,\Omega ]$ ($[-\Delta ,\Delta ]$), $\Omega $ and $\Delta $ being free
parameters. In earlier work~\cite{JETPLett,Yuan2007}, we found that a
Heisenberg-like $H_{e}$ can induce close to perfect decoherence of the
central system and therefore, we will focus on this case only.

The bath is further characterized by the number of environment spins $K$
with which a spin in the environment interacts. If $K=0$, each spin in the
environment interacts with the central system only. $K=2$, $K=4$ or $K=6$
correspond to environments in which the spins are placed on a ring, square
or triangular lattice, respectively and interact with nearest-neighbors
only. If $K=N-1$, each spin in the environment interacts with all the other
spins in the environment and, to give this case a name, we will refer to
this case as \textquotedblleft spin glass\textquotedblright .

If the Hamiltonian of the central system $H_{c}$ is a perturbation relative
to the interaction Hamiltonian $H_{ce}$, the pointer states are eigenstates
of $H_{ce}$~\cite{zurek}. In the opposite case, that is the regime $|\Delta
|\ll |J|$ that we explore in this paper, the pointer states are conjectured
to be eigenstates of $H_{c}$~\cite{paz}. The latter are given by $|1\rangle
\equiv |T_{1}\rangle =\left\vert \uparrow \uparrow \right\rangle $, $%
|2\rangle \equiv |S\rangle =(\left\vert \uparrow \downarrow \right\rangle
-\left\vert \downarrow \uparrow \right\rangle )/\sqrt{2}$, $|3\rangle \equiv
|T_{0}\rangle =(\left\vert \uparrow \downarrow \right\rangle +\left\vert
\downarrow \uparrow \right\rangle )/\sqrt{2}$, and $|4\rangle \equiv
|T_{-1}\rangle =\left\vert \downarrow \downarrow \right\rangle $, satisfying
$H_{c}|S\rangle =(3J/4)|S\rangle $ and $H_{c}|T_{i}\rangle
=(-J/4)|T_{i}\rangle $ for $i=-1,0,1$.

The simulation procedure is as follows. We generate a random superposition $%
\left\vert \phi \right\rangle $ of all the basis states of the environment.
This state corresponds to the equilibrium density matrix of the environment
at infinite temperature.
Alternatively, to study the effect of the thermal state of the environment
on the decoherence processes, we take the 2 state of
the environment to be its ground state.
The spin-up -- spin-down state ($\left\vert
\uparrow \downarrow \right\rangle $) is taken as the initial state of the
central system. Thus, the initial state of the whole system reads $%
\left\vert \Psi (t=0)\right\rangle =\left\vert \uparrow \downarrow
\right\rangle \left\vert \phi \right\rangle $ and is a product state of the
state of the central system and the initial state of the environment which,
in general is a (very complicated) linear combination of the $2^{N}$ basis
states of the environment. In our simulations we take $N=16$ which, from
earlier work~\cite{JETPLett,Yuan2007}, is sufficiently large for the
environment to behave as a \textquotedblleft large\textquotedblright\
system.

For a given, fixed set of model parameters, the time evolution of
the whole system is obtained by solving the time-dependent Schr\"{o}dinger
equation 
for the many-body wave function $|\Psi (t)\rangle $, describing the central
system plus the environment~\cite{method}. It conserves the energy of the
whole system to machine precision. We monitor the effects of the decoherence
by computing the the matrix elements of the reduced density matrix $\rho
\left( t\right) $ of the central system.

As explained earlier, in the regime
of interest $|\Delta |\ll |J|$, the pointer states are expected to be the
eigenstates of the central systems. Hence we compute the matrix elements of
the density matrix in the basis of eigenvectors of the central system. We
also compute the time dependence of quadratic entropy $S_{c}\left( t\right)
=1-Tr\rho ^{2}\left( t\right) $ and the Loschmidt echo $L\left( t\right)
=Tr\left( \rho \left( t\right) \rho _{0}\left( t\right) \right) $~\cite%
{Cucchietti2003}, where $\rho _{0}\left( t\right) $ is the density matrix
for $H_{ce}=0$.

\begin{figure}[t]
\begin{center}
\includegraphics[width=9cm]{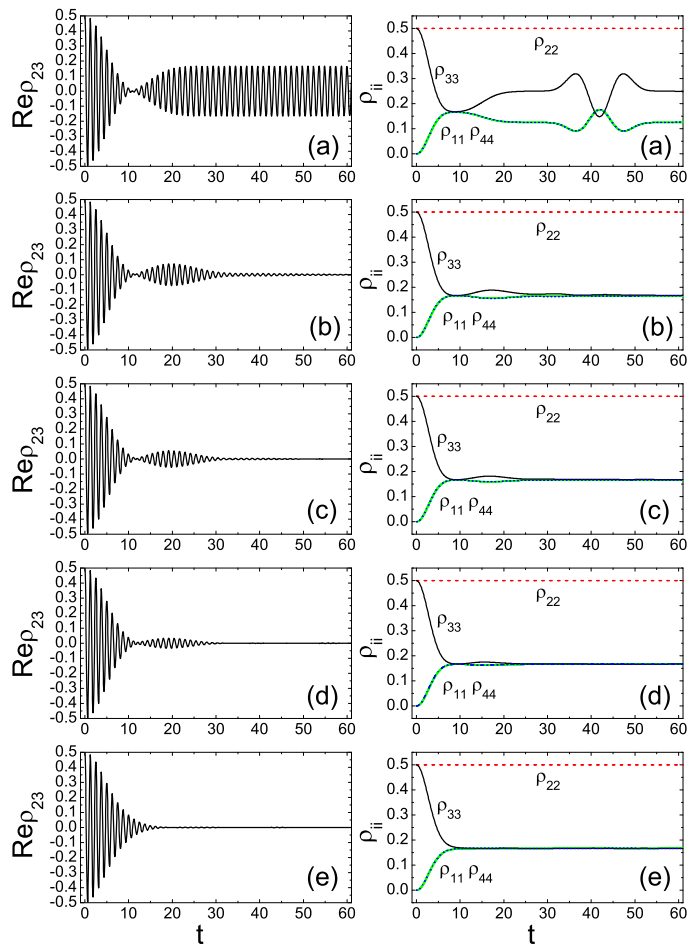}
\end{center}
\caption{(Color online) The time evolution of the real part of the
off-diagonal element $\protect\rho _{23}$ (left panel) and the diagonal
elements $\protect\rho _{11},\ldots ,\protect\rho _{44}$ (right panel) of
the reduced density matrix of a central system (with $J=-5$), coupled via an
isotropic Heisenberg interaction $H_{ce}$ ($\Delta =-0.075$ ) to a
Heisenberg-like environment $H_{e}$ ($\Omega =0.15$) with different
connectivity: (a) $K=0$; (b) $K=2$; (c) $K=4$; (d) $K=6$; (e) $K=N-1$. }
\label{fig1}
\end{figure}

\begin{figure*}[t]
\begin{center}
\includegraphics[width=12.5cm]{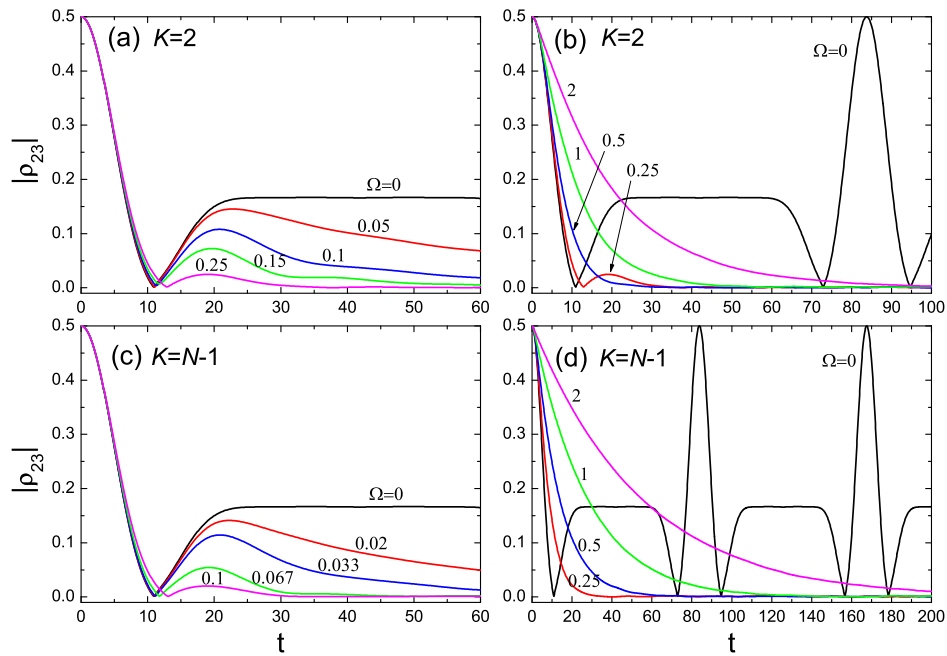}
\end{center}
\caption{(Color online) The time evolution of the off-diagonal element $%
\protect\rho _{23}$ of the reduced density matrix of a central system (with $%
J=-5$), interacting with a Heisenberg-like environment $H_{e}$ via an
isotropic Heisenberg Hamiltonian $H_{ce}$ (with $\Delta =-0.075$ ) for the
same geometric structures in the environment: (a,b) $K=2$ and (c,d) $K=N-1$.
The number next to each curve is the corresponding value of $\Omega $.}
\label{fig2}
\end{figure*}

\section{Isotropic Coupling to the Bath}

If the interaction between the central system and environment is isotropic
we have $[H_{c},H_{ce}]=0$.
Then, as shown in the Appendix, the expressions of the reduced density matrix
$\rho \left( t\right) $ and the Loschmidt echo $L\left( t\right) $
simplify.
Indeed, if $\Delta _{i,j}^{(x)}=\Delta _{i,j}^{(y)}=\Delta _{i,j}^{(z)}\equiv \Delta $ for all $i,j$,
then
\begin{equation}
H_{ce}=-\Delta (\mathbf{S}_{1}+\mathbf{S}_{2})\cdot \sum_{j=1}^{N}\mathbf{I}%
_{j}  \label{Hce1}
\end{equation}
commutes with $H_{c}$ and it follows that the decoherence process of the central
system is determined by $H_{ce}$, $H_{e}$,
the initial state of whole system $\left\vert \Psi (t_{0})\right\rangle $,
and the eigenstates of the central system (see Eq.~(\ref{L1}) and (\ref{P1}) in the Appendix).
In other words, in this case, $L\left( t\right) $ and $\left\vert \rho \left( t\right) \right\vert $
do not dependent on the $J$, the interaction between the spins in the central system.
Furthermore, if we take the interactions between the environment spins to be
isotropic, that is, $\Omega _{i,j}^{(x)}=\Omega _{i,j}^{(y)}=\Omega_{i,j}^{(z)}\equiv \Omega _{i,j}$
for all $i,j$, then
\begin{equation}
H_{e}=-\sum_{i=1}^{N-1}\sum_{j=i+1}^{N}\Omega _{i,j}\mathbf{I}_{i}\cdot
\mathbf{I}_{j}  \label{He1}
\end{equation}
commutes with $H_{ce}$, and therefore $H_{e}$
has no effect on the decoherence process (see Eq.~(\ref{Phi2}) in the Appendix).

In Fig.~\ref{fig1} and Fig.~\ref{fig2}, we show the time evolution of the
elements of the reduced density matrix $\rho \left( t\right) $ for different
connectivity $K$ and $\Omega $, for the case that $H_{ce}$ is an isotropic
Heisenberg model, i.e., $\Delta _{i,j}^{(x)}=\Delta _{i,j}^{(y)}=\Delta
_{i,j}^{(z)}\equiv \Delta $ for all $i,j$.

If $\left\vert \Delta \right\vert \gg \sqrt{K}\Omega $, in agreement with
earlier work~\cite{ourPRL,Melikidze2004}, we find that in the absence of
interactions between the environment spins ($\sqrt{K}\Omega =0$) and after
the initial decay, the central system exhibits long-time oscillations (see
Fig.~\ref{fig1}(a)(left)). In this case and in the limit of a large
environment, we have~\cite{Melikidze2004}
\begin{equation}
\hbox{Re }\rho _{23}\left( t\right) =\left[ \frac{1}{6}+\frac{1-bt^{2}}{3}%
e^{-ct^{2}}\right] \cos \omega t,  \label{melik}
\end{equation}%
where $b=N\Delta ^{2}/4$, $c=b/2$ and $\omega =J-\Delta $. Equation~(\ref%
{melik}) clearly shows the two-step process, that is, after the initial
Gaussian decay of the amplitude of the oscillations, the oscillations revive
and their amplitude levels of~\cite{Melikidze2004}. Due to conservation
laws, this behavior does not change if the environment consists of an
isotropic Heisenberg system ($\Omega _{i,j}^{(\alpha )}\equiv \Omega $ for
all $\alpha $, $i$ and $j$), independent of $K$. If, as in Ref.~\cite{ourPRL}%
, we take $\Delta _{i,j}^{(x)}=\Delta _{i,j}^{(y)}=\Delta _{i,j}^{(z)}\in %
\left[ 0,\Delta \right] $ random instead of the identical, the amplitude of
the long-living oscillations is no longer constant but decays very slowly~%
\cite{ourPRL} (results not shown).

If $\left\vert \Delta \right\vert \approx \sqrt{K}\Omega $, the presence of
Heisenberg-like interactions between the spins of the environment has little
effect on the initial Gaussian decay of the central system, but it leads to
a reduction and to a decay of the amplitude of the long-living oscillations.
The larger $K$ (see Fig.~\ref{fig1}(b-e)(left)) or $\Omega $ ({see Fig.~\ref%
{fig2}(a,c)}), the faster the decay is. Note that for the sake of clarity,
we have suppressed the fast oscillations by plotting instead of the real
part, the absolute value of the matrix elements.

If $\left\vert \Delta \right\vert \ll \sqrt{K}\Omega $, keeping $K$ fixed
and increasing $\Omega $ smoothly changes the initial decay from Gaussian
(fast) to exponential (slow), and the long-living oscillations are
completely suppressed ({see Fig.~\ref{fig2}(b,d)}). For large $\Omega $, the
simulation data fits very well to
\begin{equation}
\left\vert \rho _{23}\left( t\right) \right\vert =\frac{1}{2}e^{-A_{K}\left(
\Omega \right) t},  \label{p23_exp}
\end{equation}%
with $A_{K}\left( \Omega \right) \approx \Omega \widetilde{A}_{K}$, $%
\widetilde{A}_{2}=9.13$ and $\widetilde{A}_{N-1}=26.73$.
Note that, in principle, a closed quantum system cannot exhibit exponential decay~\cite{BALL03}.
The fact that we observe a decay that is well described by a single exponential
may be the result of tracing out the degrees of freedom of an environment which initially is
in a state of random superposition of the basis states.

\begin{figure*}[t]
\begin{center}
\includegraphics[width=12.5cm]{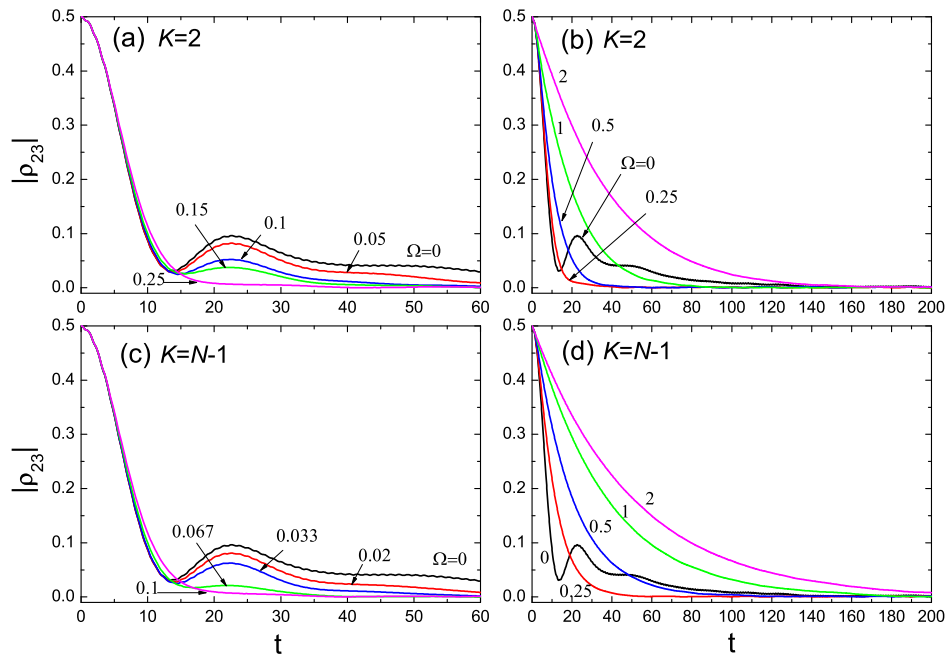}
\end{center}
\caption{(Color online) Same as Fig.~\protect\ref{fig2} except that $H_{ce}$
is Heisenberg-like and $\Delta =0.15$. }
\label{fig3}
\end{figure*}

Physically, the observed behavior can be understood as follows. If $%
\left\vert \Delta \right\vert \approx \sqrt{K}\Omega $, a bath spin is
affected by roughly the same amount by the motion of both the other bath
spins and by the two central spins. Therefore, each bath spin has enough
freedom to follow the original dynamics, much as if there were no coupling
between bath spins. This explains why the initial Gaussian decay is
insensitive to the values of $K$ or $\Omega $. After the initial decay, the
whole system is expected to reach an stationary state, but because of the
presence of Heisenberg-like interactions between the bath spins, a new
stationary state of the bath is established, suppressing the long-living
oscillations.

For increasing $K$, the distance between two bath spins, defined as the
minimum number of bonds connecting the two spins, becomes smaller. For
instance, for $K=2$, this distance is $\left( N-2\right) /2$, and for $K=N-1$, it is zero.
Therefore, for fixed $\Omega $ and increasing $K$ the
fluctuations in the spin bath can propagate faster and the evolution to the
stationary state will be faster. Similarly, for fixed $K$, increasing the
coupling strength between the bath spins will speed up the dynamics of the
bath, that is, the larger $\Omega $ the faster will be the evolution to the
stationary state.

In the opposite case $\left\vert \Delta \right\vert \ll \sqrt{K}\Omega $, $%
H_{ce}$ is a small perturbation relative to $H_{e}$ and the coupling between
bath spins is the dominant factor in determining the dynamics of the bath
spins. Therefore, by increasing $K$ or $\Omega $, the bath spin will have
less freedom to follow the dynamics induced by the coupling to the two
central spins, the influence of the bath on the central system will
decrease, and the (exponential) decay will become slower.

According to the general picture of decoherence~\cite{zurek}, for an
environment with nontrivial internal dynamics that initially is in a random
superposition of all its eigenstates, we expect that the central system will
evolve to a stable mixture of its eigenstates. In other words, the
decoherence will cause all the off-diagonal elements of the reduced density
matrix to vanish with time. In the case of an isotropic Heisenberg coupling
between the central system and the environment, $H_{c}$ commutes with the
Hamiltonian $H$, hence the energy of the central system is a conserved
quantity. Therefore, the weight of the singlet $\left\vert {S}\right\rangle $
in the mixed state should be a constant ($1/2$), and the weights of the
degenerate eigenstates $|T_{0}\rangle $, $|T_{-1}\rangle $ and $%
|T_{1}\rangle $ are expected to become the same (${1/6}$). As shown in Fig.~%
\ref{fig1}(b-e)(right), our simulations confirm that this picture is correct
in all respects.

\section{Anisotropic Coupling to the Bath}

In order to clarify the role of $K$ and $\Omega $, we change the coupling
between the central system and the bath from Heisenberg to Heisenberg-like.
From a comparison of the data in Fig.~\ref{fig2} and Fig.~\ref{fig3}, it is
clear that the roles of $K$ and $\Omega $ are the same in both cases, no
matter whether the coupling to the bath is isotropic or anisotropic.
However, there are some differences in the decoherence process.
The most important parameter determining the decoherence process
is the ratio of the typical interaction energy $\Delta$
to the mean-square energy of interactions in the the thermal bath, $\sqrt{K}\Omega$.

If $\left\vert \Delta \right\vert \gg \sqrt{K}\Omega $, in the presence of
anisotropic interactions between the central system and the environment
spins, even in the absence of interactions between the bath spins, the
second step of the oscillations decays and finally disappear as $K$
increases. This is because the anisotropic interactions break the rotational
symmetry of the coupling between central system and environment which is
required for the long-living oscillations to persist.

If $\left\vert \Delta \right\vert \ll \sqrt{K}\Omega $, $\left\vert \rho
_{23}\left( t\right) \right\vert $ can still be described by Eq.~(\ref%
{p23_exp}), but now $A_{K}\left( \Omega \right) $ is no longer a linear
function of $\Omega $. For anisotropic $H_{ce}$, the energy of the
central system is no longer a conserved quantity. Therefore there will be
energy transfer between the central system and the environment and the
weight of each pointer state (eigenstate) in the final stable mixture need
not be the same for all $K$ or $\Omega $.

\begin{figure}[t]
\begin{center}
\includegraphics[width=8.5cm]{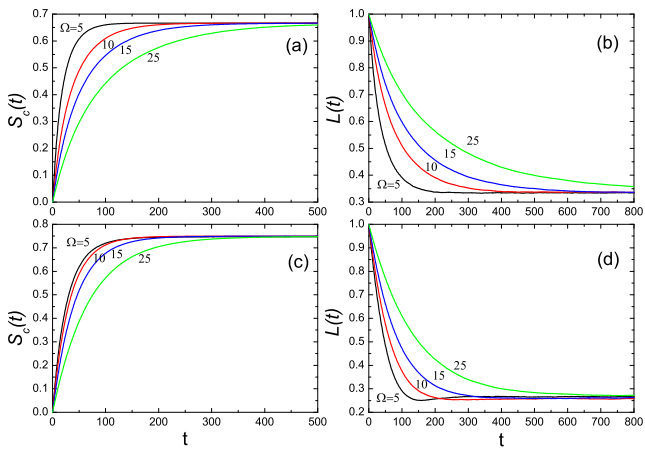}
\end{center}
\caption{(Color online) The time evolution of the the entropy $S_{c}\left(
t\right) $ and Loschmidt echo $L\left( t\right) $ of a central system (with $%
J=-5$), interacting with a Heisenberg-like environment $H_{e}$ (with
different $\Omega $) via a Heisenberg (a,b, $\Delta =-0.075$) or
Heisenberg-like (c,d, $\Delta =0.15$) Hamiltonian $H_{ce}$ for the case $K=2$.
The number next to each curve is the corresponding value of $\Omega $.}
\label{fig4}
\end{figure}

For a change, we illustrate this point by considering the quadratic entropy $%
S_{c}\left( t\right) $ and Loschmidt echo $L\left( t\right) $. We expect
that these quantities will also dependent of the symmetry of the coupling
between central system and the spin bath. In Fig.~\ref{fig4}, we present
results for large $\Omega $ and $K=2$, confirming this expectation. For
isotropic (Heisenberg) $H_{ce}$ and perfect decoherence (zero off-diagonal
terms in the reduced density matrix) we expect that $\max_{t}S_{c}(t)=1-[%
\left( 1/2\right) ^{2}+3\times \left( {1/6}\right) ^{2}]=2/3$, in concert
with the data of Fig.~\ref{fig4}(a)). For Heisenberg-like $H_{ce}$, $%
\max_{t}S_{c}(t)$ will depend on the coupling strengths and as shown in Fig.~%
\ref{fig4}(c), we find that $\max_{t}S_{c}(t)=$ $1-4\times \left( {1/4}%
\right) ^{2}=3/4$, corresponding to the case that all the diagonal elements
in the reduced density matrix are the same ($1/4$) and all other elements
are zero.

\begin{figure}[t]
\begin{center}
\includegraphics[width=8.5cm]{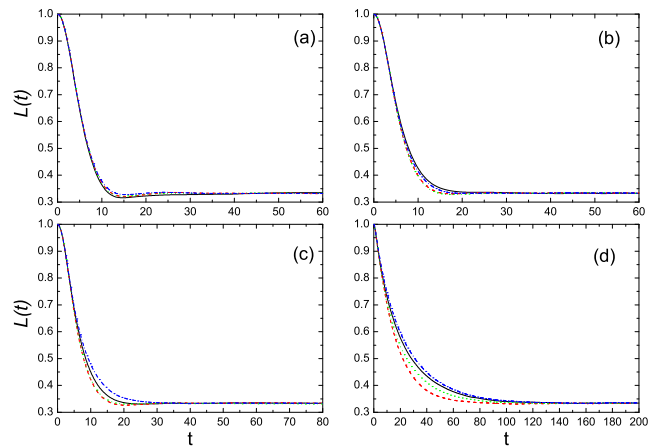}
\end{center}
\caption{(Color online) The time evolution of the Loschmidt echo $L\left(
t\right) $ of a central system (with $J=-5$), interacting with a
Heisenberg-like environment $H_{e}$ via a Heisenberg ($\Delta =-0.075$)
Hamiltonian $H_{ce}$. In each panel, the values of $\protect\sqrt{K}\Omega $
are the same: (a) $\protect\sqrt{K}\Omega \equiv 0.1\protect\sqrt{N-1}$, (b)
$\protect\sqrt{K}\Omega \equiv 0.15\protect\sqrt{N-1}$, (c) $\protect\sqrt{K}%
\Omega \equiv 0.25\protect\sqrt{N-1}$, and (d) $\protect\sqrt{K}\Omega
\equiv \protect\sqrt{N-1}$. The different lines in each pannel correspond to
different $K$.
Solid (black) line: $K=2$; Dashed (red) line: $K=4$; Dotted (green) line:
$K=6$, and dash-dotted (blue) line: $K=N-1$.}
\label{fig5}
\end{figure}

\begin{figure}[t]
\begin{center}
\includegraphics[width=8cm]{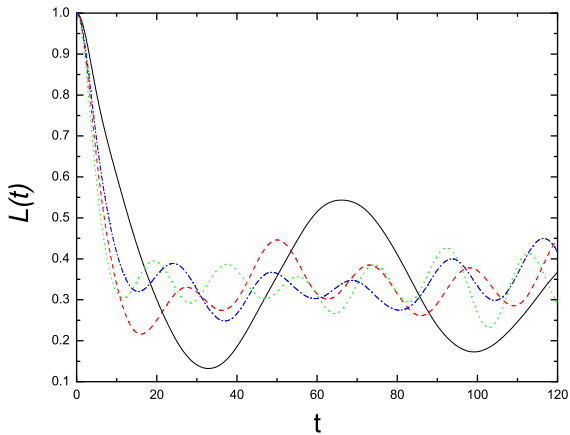}
\end{center}
\caption{(Color online) The time evolution of the Loschmidt echo $L\left(
t\right) $ of a central system (with $J=-5$), interacting with a
Heisenberg-like environment $H_{e}$ via a Heisenberg ($\Delta =-0.075$)
Hamiltonian $H_{ce}$. The environment spins are initially prepared in the
ground state.
The different curves correspond to different $K$, but $\protect\sqrt{K}\Omega=0.15\protect\sqrt{N-1}$ is fixed.
Solid (black) line: $K=2$; Dashed (red) line: $K=4$; Dotted (green) line:
$K=6$, and dash-dotted (blue) line: $K=N-1$.}
\label{fig6}
\end{figure}

\section{Discussion and conclusions}

In the foregoing,
we have compared $\sqrt{K}\Omega $ to $\left\vert \Delta \right\vert $
to distinguish different regimes.
As a matter of fact, $\sqrt{K}\Omega $
does not completely characterize the decoherence process,
but it can be used to characterize its time scale.
Indeed, as shown in Fig.~\ref{fig5}, for different $\sqrt{K}$ and $\Omega $
but the same value of $\sqrt{K}\Omega $, the
the time evolution of $L(t)$ is very similar.
Note that if $\sqrt{K}\Omega $ increases (compare Fig.~\ref{fig5}a to  Fig.~\ref{fig5}d),
the differences between the Loschmidt echoes increase.
Additional simulations (results not shown) indicate
that this differences are fluctuations that are due
to the particular realization random parameters used in the simulation.

In conclusion, for a spin-bath environment that initially is
in a random superposition of its basis states,
we have shown how a pure quantum state of the central spin
system evolves into a mixed state, and that if the interaction between the
central system and environment is much smaller than the coupling between the
spins in the central system, the pointer states are the eigenstates of the
central system. Both these observations are in concert with the general
picture of decoherence~\cite{zurek}. Furthermore, we have demonstrated that,
in the case that the environment is a spin system, the details of this spin
system are important for the decoherence of the central system. In
particular, we have shown that for the anisotropic spin-bath, changing the
internal dynamics of the environment (geometric structure or exchange
couplings) may change the decoherence of the central spin system from
Gaussian to exponential decay.

Finally, we would like to compare the present results with those
of our earlier work in which we focussed on the case
in which the environment is initially in its ground state and
demonstrated that, apart from the strength of different
interactions, also their symmetry and the amount of entanglement
of the ground state of the central system affects the decoherence~\cite{JETPLett,Yuan2007}.
To facilitate the comparison, in Fig.~\ref{fig6}
we present some new data of the Loschmidt echoes for different $K$
but for fixed $\protect\sqrt{K}\Omega$.
Comparison of Fig.~\ref{fig5} with Fig.~\ref{fig6} indicates
that if the environment is initially in its ground state,
the decoherence process is qualitatively different from
the one observed in the case that the initial state
of the environment is a random superposition.
Roughly speaking, it is more difficult for the central system to change
from a pure quantum state to a classical, mixed state,
which is of course consistent with the fact that the quantum effects become
more prominent as the temperature decreases.
In particular, from Fig.~\ref{fig6} it is clear that
$\sqrt{K}\Omega $ is not enough to characterize
the qualitative behavior of the Loschmidt echo for the cases shown.

The difference between the cases of an environment at low-temperature~\cite{JETPLett,Yuan2007}
and a high-temperature (chaotic) environment considered in the present paper is
most important for the systems with very large connectivity.
In the latter case, the ground state of the environment is a quantum spin-glass which is a very effective source of decoherence~\cite{JETPLett,Yuan2007}.
At the same time, for the case of infinite temperature of the bath considered in this paper,
this case is not very special when compared to the case of short-range interactions within the environment (see Fig.~\ref{fig5}).
It would be of interest to see if, as the temperature decreases, the decoherence process changes
as the environment goes into the spin-glass state (at $T \propto \sqrt{K}\Omega$), a problem that we leave for future research.

\section{Appendix}

Consider a generic quantum model described by the Hamiltonian $%
H=H_{c}+H_{ce}+H_{e}$, where $H_{c}$ and $H_{e}$ describe the central system
and the bath respectively ($[H_{c},H_{e}]=0$), and $H_{ce}$ describes the
coupling between them. If $[H_{c},H_{ce}]=0$, then the time evolution
operator of the whole system $e^{-iHt}$ can be represented as%
\begin{equation}
e^{-iHt}=e^{-iH_{c}t}e^{-i\left( H_{ce}+H_{e}\right) t}.
\end{equation}%
Denote the eigenstates and corresponding eigenvalues of the central system
by $\{\left\vert k\right\rangle \}$ and $\{E_{k}\}$, that is, $%
H_{c}\left\vert k\right\rangle =E_{k}\left\vert k\right\rangle $.
The initial state ($\left\vert \varphi (t_{0})\right\rangle $) of the central
system can be represented as $\left\vert \varphi (t_{0})\right\rangle
=\sum_{k}a_{k}\left\vert k\right\rangle $.
For an isolated central system ($H_{ce}=0$),
the time evolution of the density matrix of the central system is given by
\begin{equation}
\rho _{0}\left( t\right) =\sum_{k,l}e^{-i\left( E_{k}-E_{l}\right)
t}a_{k}a_{l}^{\ast }\left\vert k\right\rangle \left\langle l\right\vert .
\end{equation}%
If the central system is coupled to the bath ($\left\vert \phi \left(
t_{0}\right) \right\rangle $), the initial state of the whole system can be
represent as
\begin{equation}
\left\vert \Psi (t_{0})\right\rangle =\sum_{k}a_{k}\left\vert k\right\rangle
\left\vert \phi \left( t_{0}\right) \right\rangle ,
\end{equation}%
and the state at later time $t$ is
\begin{eqnarray}
\left\vert \Psi (t)\right\rangle \rangle  &=&e^{-iHt}\left\vert \Psi
(t_{0})\right\rangle   \notag \\
&=&\sum_{k}e^{-iE_{k}t}a_{k}e^{-i\left( H_{ce}+H_{e}\right) t}\left\vert
k\right\rangle \left\vert \phi \left( t_{0}\right) \right\rangle .
\end{eqnarray}%
As $[H_{c},H_{ce}]=0$, we have $H_{ce}\left\vert k\right\rangle \left\vert
\phi \left( t_{0}\right) \right\rangle =\left\vert k\right\rangle
M_{k}\left\vert \phi \left( t_{0}\right) \right\rangle $, therefore
\begin{eqnarray}
&&e^{-i\left( H_{ce}+H_{e}\right) t}\left\vert k\right\rangle \left\vert
\phi \left( t_{0}\right) \right\rangle   \notag \\
&=&\sum_{m}\frac{\left( -it\right) ^{m}\left( H_{ce}+H_{e}\right) ^{m}}{m!}%
\left\vert k\right\rangle \left\vert \phi \left( t_{0}\right) \right\rangle
\notag \\
&=&\sum_{m}\left\vert k\right\rangle \frac{\left( -it\right) ^{m}\left(
M_{k}+H_{e}\right) ^{m}}{m!}\left\vert \phi \left( t_{0}\right)
\right\rangle   \notag \\
&=&\left\vert k\right\rangle e^{-i\left( M_{k}+H_{e}\right) t}\left\vert
\phi \left( t_{0}\right) \right\rangle   \notag \\
&=&\left\vert k\right\rangle \left\vert \phi _{k}\left( t\right)
\right\rangle ,
\end{eqnarray}%
where we introduced
\begin{equation}
\left\vert \phi _{k}\left( t\right) \right\rangle \equiv e^{-i\left(
M_{k}+H_{e}\right) t}\left\vert \phi \left( t_{0}\right) \right\rangle .
\label{phit}
\end{equation}%
Hence, the state at time $t$ becomes%
\begin{equation}
\left\vert \Psi (t)\right\rangle =\sum_{k}a_{k}e^{-iE_{k}t}\left\vert
k\right\rangle \left\vert \phi _{k}\left( t\right) \right\rangle .
\end{equation}%
The density matrix $\rho \left( t\right) $ of the whole system is%
\begin{eqnarray}
\rho \left( t\right)  &=&\left\vert \Psi (t)\right\rangle \left\langle \Psi
(t)\right\vert   \notag \\
&=&\sum_{k,l}e^{-i\left( E_{k}-E_{l}\right) t}a_{k}a_{l}^{\ast }\left\vert
k\right\rangle \left\vert \phi _{k}\left( t\right) \right\rangle
\left\langle l\right\vert \left\langle \phi _{l}\left( t\right) \right\vert ,
\notag \\
&&
\end{eqnarray}%
and the reduced density matrix $\rho _{c}\left( t\right) $ of the central
system is%
\begin{eqnarray}
\rho _{c}\left( t\right)  &=&Tr_{e}\rho \left( t\right)   \notag \\
&=&\sum_{k,l}e^{-i\left( E_{k}-E_{l}\right) t}a_{k}a_{l}^{\ast }\left\langle
\phi _{l}\left( t\right) |\phi _{k}\left( t\right) \right\rangle \left\vert
k\right\rangle \left\langle l\right\vert .  \notag \\
&&
\end{eqnarray}%
The Loschmidt echo $L\left( t\right) $ of the central system can be
calculated as
\begin{eqnarray}
L\left( t\right)  &=&Tr\left( \rho _{c}\left( t\right) \rho _{0}\left(
t\right) \right)   \notag  \label{L1} \\
&=&Tr{ [}\sum_{k,l}e^{-i\left( E_{k}-E_{l}\right) t}a_{k}a_{l}^{\ast
}\left\langle \phi _{l}\left( t\right) |\phi _{k}\left( t\right)
\right\rangle \left\vert k\right\rangle \left\langle l\right\vert   \notag \\
&&\times \sum_{m,n}e^{-i\left( E_{m}-E_{n}\right) t}a_{m}a_{n}^{\ast
}\left\vert m\right\rangle \left\langle n\right\vert { ]}  \notag \\
&=&Tr{ [}\sum_{k,l,n}e^{-i\left( E_{k}-E_{n}\right) t}a_{k}\left\vert
a_{l}\right\vert ^{2}a_{n}^{\ast }  \notag \\
&&\times \left\langle \phi _{l}\left( t\right) |\phi _{k}\left( t\right)
\right\rangle _{l}\left\vert k\right\rangle \left\langle n\right\vert { %
]}  \notag \\
&=&\sum_{k,l}\left\vert a_{k}\right\vert ^{2}\left\vert a_{l}\right\vert
^{2}\left\langle \phi _{l}\left( t\right) |\phi _{k}\left( t\right)
\right\rangle .
\end{eqnarray}

It is clear that if $[H_{c},H_{ce}]=0$, the decoherence process is
determined by the initial state of the central system $\{a_{k}\}$ and the
time evolution of the $\{\left\vert \phi _{k}\left( t\right) \right\rangle
\} $. As shown in Eq.~(\ref{phit}), the $\{\left\vert \phi _{k}\left( t\right)
\right\rangle \}$ are determined by the initial state of the bath $\left(
\left\vert \phi \left( t_{0}\right) \right\rangle \right) $, the eigenstates
$\{\left\vert k\right\rangle \}$ of the central system, and the Hamiltonian $%
H_{ce}$ and $H_{e}$. The eigenvalues $\{E_{k}\}$ have no effect of the
decoherence process. Thus, multiplying $H_{c}$ by a constant does not
change the $L\left( t\right) $ and the diagonal elements of the
reduced density matrix $\rho _{c}\left( t\right) $.
The time evolution of the absolute value of the off-diagonal elements
\begin{equation}
\left\vert \rho _{c}\left( t\right) _{kl}\right\vert =\left\vert
a_{k}a_{l}^{\ast }\right\vert \left\langle \phi _{l}\left( t\right) |\phi
_{k}\left( t\right) \right\rangle,  \label{P1}
\end{equation}%
is independent of $H_{c}$.

Finally, we consider the case that not only $[H_{c},H_{ce}]=0$ but also $%
[H_{ce},H_{e}]=0$. Then, Eq.~(\ref{phit}) becomes%
\begin{equation}
\left\vert \phi _{k}\left( t\right) \right\rangle =e^{-i\left(
M_{k}+H_{e}\right) t}\left\vert \phi \left( t_{0}\right) \right\rangle
=e^{-iM_{k}t}e^{-iH_{e}t}\left\vert \phi \left( t_{0}\right) \right\rangle ,
\end{equation}%
therefore we have%
\begin{eqnarray}
\left\langle \phi _{l}\left( t\right) |\phi _{k}\left( t\right)
\right\rangle &=&\left\langle \phi \left( t_{0}\right) \right\vert
e^{iH_{e}t}e^{iM_{l}t}e^{-iM_{k}t}e^{-iH_{e}t}\left\vert \phi \left(
t_{0}\right) \right\rangle  \notag  \label{Phi2} \\
&=&\left\langle \phi \left( t_{0}\right) \right\vert e^{-i\left(
M_{k}-M_{l}\right) t}\left\vert \phi \left( t_{0}\right) \right\rangle ,
\end{eqnarray}%
implying that $\left\vert \rho _{c}\left( t\right) _{kl}\right\vert $ and $%
L\left( t\right) $ do not dependent on $H_{e}$.

\section*{Acknowledgement}

M.I.K. acknowledges a support by the Stichting Fundamenteel Onderzoek der
Materie (FOM).

\end{document}